# 3-D Integrated Flexible Glass Photonics

Lan Li[1†], Hongtao Lin[1†], Shutao Qiao[2†], Yi Zou[1], Sylvain Danto[3], Kathleen Richardson[3, 4], J. David Musgraves[4], Nanshu Lu[2], and Juejun Hu[1*]

*1. Department of Materials Science and Engineering*

*University of Delaware*

*Newark, Delaware 19716*

*2. Center for Mechanics of Solids, Structures and Materials*

*Department of Aerospace Engineering and Engineering Mechanics*

*University of Texas at Austin*

*Austin, Texas 78712*

*3. College of Optics and Photonics, CREOL*

*Department of Materials Science and Engineering*

*University of Central Florida*

*Orlando, Florida 32816*

*4. IRradiance Glass Inc.*

*Orlando, FL 32828, USA*

†These authors contributed equally to this work.

*e-mail: hujuejun@udel.edu

**Abstract:**

Photonic integration on plastic substrates enables emerging applications ranging from flexible interconnects to conformal sensors on biological tissues. Such devices are traditionally fabricated using pattern transfer, which is complicated and has limited integration capacity. Here we pioneered a monolithic approach to realize flexible, high-index-contrast glass photonics with significantly improved processing throughput and yield. Noting that the conventional multilayer bending theory fails when laminates have large elastic mismatch, we derived a mechanics theory accounting for multiple neutral axes in one laminated structure to accurately predict its strain-optical coupling behavior. Through combining monolithic fabrication and local neutral axis designs, we fabricated devices that boast record optical performance (Q=460,000) and excellent mechanical flexibility enabling repeated bending down to sub-millimeter radius without measurable performance degradation, both of which represent major improvements over state-of-the-art. Further, we demonstrate that our technology offers a facile fabrication route for 3-D high-index-contrast photonics difficult to process using traditional methods.

While conventional on-chip photonic devices are almost exclusively fabricated on rigid substrates with little mechanical flexibility, integration on deformable polymer substrates has given birth to flexible photonics, a field which has rapidly emerged in recent years to the forefront of photonics. By imparting mechanical flexibility to planar photonic structures, the technology has seen enormous application potential for aberration-free optical imaging[1], epidermal sensing[2], chip-to-chip interconnects[3], and broadband photonic tuning[4]. Free-space-coupled optical components including photodetectors[5], light emitting diodes[6], and Fano reflectors[7] are among the first flexible semiconductor photonic devices demonstrated. Planar integrated photonic structures such as micro-resonators and other waveguide-integrated devices potentially offer significantly improved performance characteristics compared to their free-space counterparts given their tight optical confinement, while also facilitating essential components for integrated photonic circuits. To date, flexible planar photonic devices were almost exclusively fabricated using polymeric materials, which do not possess the high refractive indices necessary for strong optical confinement. Silicon-based, high-index-contrast flexible waveguide devices were first demonstrated using a transfer printing approach[8]. This hybrid approach, however, involves multiple pattern transfer steps between different substrates and has limited throughput, yield, and integration capacity[9,10]. More recently, amorphous silicon devices were fabricated via direct deposition and patterning[11]. However, the silicon material optical quality was severely compromised by the low deposition temperature dictated by the polymer substrate's thermal budget.

Besides these multi-material integration challenges, planar integrated photonics stipulate a distinctively different set of requirements on the configurational design to attain structural flexibility. For example, the neutral plane design widely adopted for flexible electronics[12,13] dictates that the device layer should be embedded inside the flexible substrate near the neutral plane to minimize strain exerted on the devices when the structure is deformed. However, encapsulation of the various photonic components deep within a thick top cladding layer prohibits effective heat dissipation as well as evanescent wave interactions with the external environment, an essential condition for biochemical sensing and evanescent optical coupling. To achieve efficient optical coupling, current flexible photonic devices are placed on the surface of polymer substrates. As a consequence, the devices is subjected to large strains upon bending and exhibits only moderate flexibility with a bending radius typically no less than 5 mm[3,5,11]. This mechanical performance severely limits possible deployment degrees of freedom for devices of this type.

In this work, we demonstrate monolithic photonic integration on plastic substrates using high-refractive index chalcogenide glass (ChG) materials. This process yields high-index-contrast nanophotonic devices with record optical performance and benefits significantly from improved processing protocols based on simple, low-cost contact lithography. We show that this versatile process can be readily adapted to different glass compositions with tailored optical properties to meet different candidate applications. A novel local-neutral-axis design is implemented to render the structure highly mechanically flexible, enabling repeated bending of the devices down to sub-millimeter bending radius without measurable optical performance degradation. We note that the classical multilayer beam bending theory fails in our design due to the large modulus contrast among different layers. For this reason a new analytical model accounting for the multiple neutral axes in a multilayer stack was developed to successfully capture the strain-optical coupling behavior in our devices. In addition, we have further exploited the monolithic integration approach for 3-D multilayer fabrication. Compared to conventional 3-D stacking methods involving wafer bonding[14], nano-manipulation[15], ion implantation[16], or multi-step chemical mechanical polishing (CMP)[17], our approach offers a simple and robust alternative route for novel 3-D photonic structure processing.

Fig. 1a schematically illustrates our device fabrication process flow. The process starts with epoxy (SU-8) spin-coating on a rigid handler (e.g. an oxide coated Si wafer), followed by ChG evaporation deposition and lift-off patterning using UV contact lithography (see Methods). We choose ChGs as the photonic materials for several reasons: their amorphous nature and low deposition temperature permit direct monolithic flexible substrate integration[18,19]; their high refractive indices (2 to 3) are compatible with high-index-contrast photonic integration; and their almost infinite capacity for composition alloying allows fine tuning of optical as well as thermal-mechanical properties over a broad range, making them suitable for diverse applications. SU-8 epoxy was used as the cladding polymer given its proven chemical stability, excellent optical transparency, and superior planarization capacity. Prior to UV exposure, SU-8 epoxy behaves as a thermoplastic polymer amenable to thermal reflow treatments to create a smooth surface finish even on substrates with multi-level patterned structures; after thermal or UV cross-linking, the epoxy becomes a thermosetting resin and is robust against mechanical deformation, humidity, and subsequent thermal processing. Capitalizing on this unique property of SU-8, we developed an ultra-thin epoxy planarization process with a high degree of planarization (DOP) critical to 3-D photonic integration. Details of the planarization process are described in Supplementary Materials Part I. The deposition/patterning/planarization steps were repeated multiple times for 3-D fabrication and no loss of DOP was observed. Lastly, the flexible samples were delaminated from the handler wafer using Kapton tape to form free-standing flexible structures. The Kapton tape consists of two layers, a silicone adhesive layer and a polyimide substrate, and it serves dual purposes: firstly, it facilitates the delamination process; and secondly, the low-modulus silicone adhesives serves as an effective strain-relieving agent in our local-neutral-axis design to be discussed later. Fig. 1b shows a photo of a final free-standing flexible photonic circuit chip. We have also tested the fabrication process with several different ChG compositions with vastly different optical properties (indices and Tauc optical band gaps) to demonstrate the process' material compatibility: Fig. 1c shows University of Delaware logos patterned on flexible substrates made of three glass compositions: $Ge_{23}Sb_7S_{70}$ ($n = 2.1$, $E_g = 2.2$ eV)[20], $As_2Se_3$ ($n = 2.8$, $E_g$ =1.8 eV)[21] and $As_2S_3$ ($n = 2.4$, $E_g = 2.1$ eV)[22], all of which exhibit good adhesion to the SU-8 substrate. The inset in Fig. 1d shows a microscopic image of a 30 μm-radius $Ge_{23}Sb_7S_{70}$ micro-disk resonator pulley coupled to a channel bus waveguide[23]. This simple fabrication route offers extremely high device yield: we have tested over 100 resonator devices randomly selected from samples fabricated in several different batches, and *all* of them operate as designed after going through the entire fabrication process. Fig. 1e plots the intrinsic Q-factor distribution of the devices measured near 1550 nm wavelength, showing an average Q-factor of $(2.7 \pm 0.7) \times 10^5$. A normalized resonator transmission spectrum is shown in Fig. 1e inset as an example: our best device exhibited an intrinsic Q-factor as high as 460,000, the highest value ever reported in photonic devices on plastic substrates.

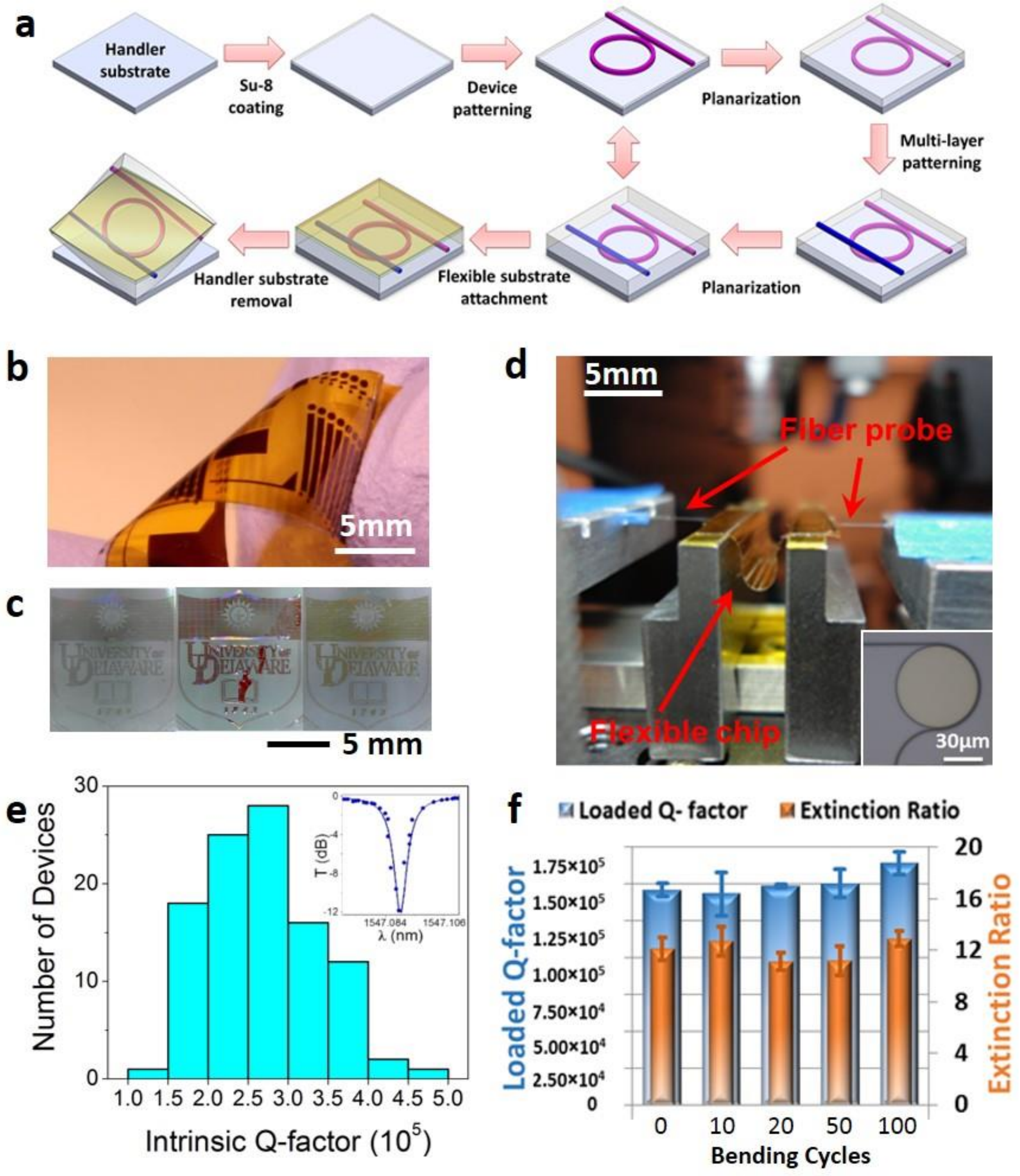

**Figure 1 | Flexible glass photonic devices fabrication and mechanical reliability tests.** **a**, Schematic overview of the monolithic 3-D flexible photonic devices fabrication process. **b**, Photo of a flexible photonic chip showing a linear array of micro-disk resonators. **c**, University of Delaware logo on PDMS flexible substrates made of $Ge_{23}Sb_7S_{70}$, $As_2Se_3$ and $As_2S_3$ glasses (from left to right). **d**, Photo of the fiber end-fire testing setup used for *in-situ* measurement of optical transmission characteristics during mechanical bending of the flexible devices. **e**, Intrinsic Q-factor distribution measured in flexible micro-disk resonators; inset shows an example of the resonator transmission spectrum. **f**, Loaded Q-factors and extinction ratios of the resonator after multiple bending cycles at 0.5 mm bending radius.

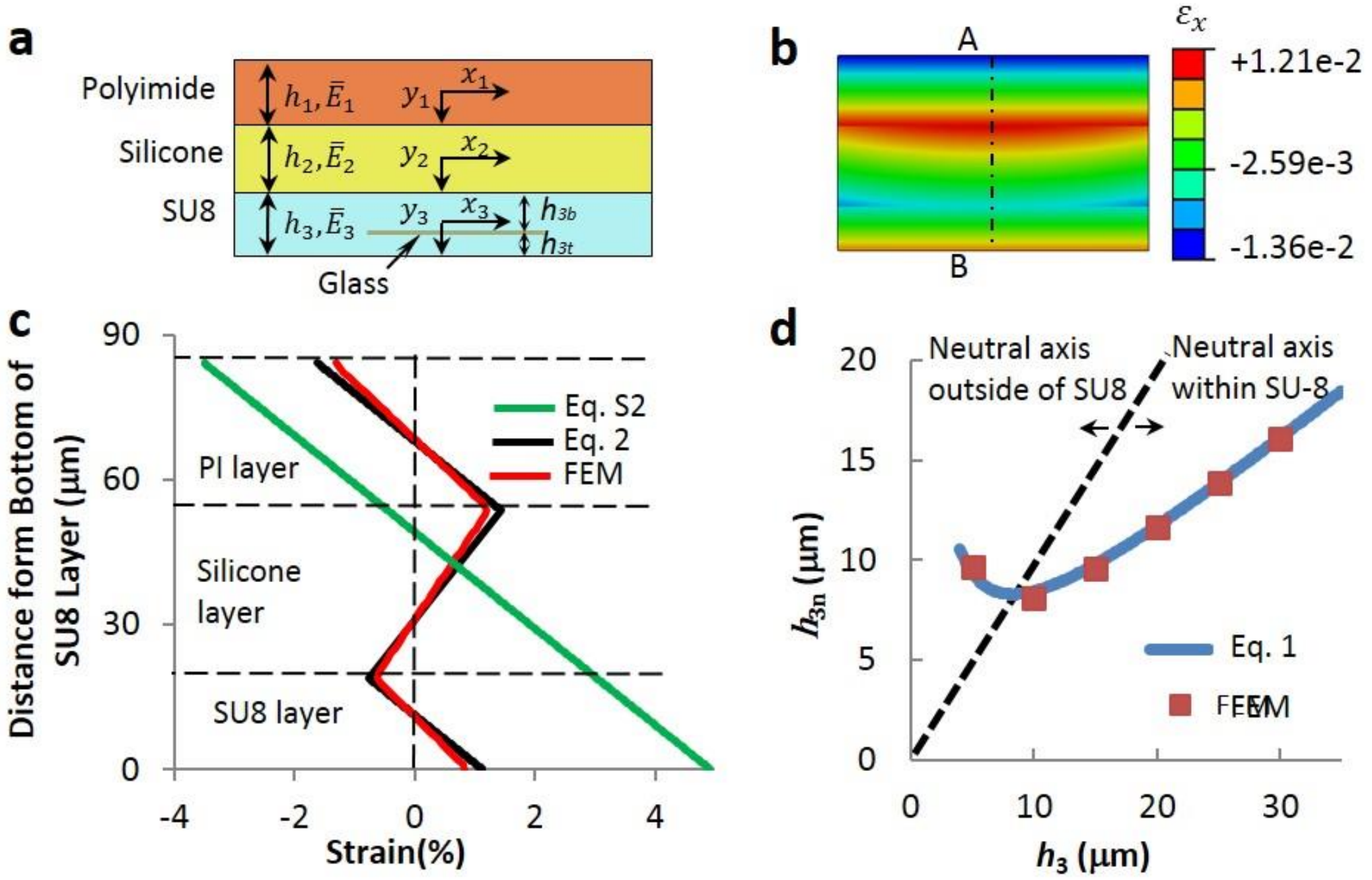

**Figure 2 | Multi-neutral-axis theory**. **a**, Cross-sectional schematic of the flexible photonic chip, with a PI-silicone-SU-8 three-layer structure (not drawn to scale), thickness and plane strain Young's modulus are as labeled, and the local coordinate originates from the median plane of each layer. **b**, Contour plots of bending strain distribution from FEM when the structure in Fig. 2a is bent. **c**, Through-stack strain distribution in the flexible chip with an SU-8 layer thickness of 18.9 μm when bent to a radius of 1 mm: the green line represents results calculated using the conventional theory (Eq. S2), the black curve is derived using our multiple neutral-axes theory (Eq. 2), and the red curve are FEM simulations. **d**, the position of SU-8 neutral axis from SU-8 surface as a function of SU-8 thickness. The blue curve comes from Eq. 1 and the solid markers are FEM results. Dash line denotes the boundary of SU-8. The neutral axis will locate outside of SU-8 to the left hand side of the dash line and hence is not accessible to the device.

Fig. 2a schematically illustrates the thickness profile of the fabricated flexible photonic chip from top to bottom comprising the polyimide (PI) substrate, the silicone adhesive, and the SU-8 cladding layer in which the devices are encapsulated. The notations are labeled in Fig. 2a and tabulated in Table S3. We experimentally measured the Young's moduli of thin film PI, silicone, and SU-8 to be $E_1$ = 2.5 GPa, $E_2$ = 1.5 MPa, and $E_3$ = 2 GPa, respectively (Supplementary Materials Part III). In the following we derive an analytical formula to predict the strain-optical coupling behavior in the devices.

The classical beam bending theory predicts that when a multilayer structure is subject to pure bending, cross-sectional planes before bending are assumed to remain planar after bending, and a unique neutral axis exists in the laminated structure (Supplementary Materials Part IV). However, the classical theory is only applicable to multilayer stacks with similar elastic stiffness. For the three-layer structure shown in Fig. 2a, the Young's modulus of silicone interlayer is three orders of magnitude smaller than that of SU-8 or PI. When this sandwiched "Oreo" structure is bent, the soft middle layer undergoes large shear deformation, which essentially decouples the deformation of the top and bottom stiff layers, similar to the strain decoupling effect discussed in the tension case[24]. As a result, each stiff layer has its own neutral axis and bending center, and thus the stack demonstrates multiple neutral axes. We first validate this postulate using finite element modeling (FEM) as shown in Fig. 2b. When the multilayer is subject to concave bending, FEM results clearly shows that the strain distribution is not monotonic across the stack thickness. Both compressive and tensile strains are present in the PI and SU-8 layers, which is contradictory to Eq. S2 where only one neutral axis exists in the structure.

To accurately predict the strain distribution in each layer, we assume that the cross-sectional planes in PI and those in SU-8 remain planar during bending. An analytical theory has been established based on continuity conditions and force equilibrium in the multilayer stack (Supplementary Materials Part IV). The distance from SU-8 surface to SU-8 neutral axis $h_{3n}$ is given by:

$$h_{3\mathrm{n}} = \frac{h_3}{2} + d_1 \frac{\bar{E}_1 h_1}{\bar{E}_3 h_3} \quad (1)$$

where $d_1$ = 0.836 μm is the distance from the PI neutral axis to the PI median plane, which is found to be a constant when silicone thickness is fixed. In a single free-standing SU-8 layer, $h_{3\mathrm{n}} = h_3/2$. When SU-8 is bonded to PI via silicone adhesive, the effect of PI is captured by the 2nd term in Eq. 1: the thicker the PI layer (i.e. the larger $h_1$), the further away the neutral axis of SU8 from its median plane.

Once the neutral axes are determined, strain distribution across thickness direction is given by

$$\varepsilon = \begin{cases} \frac{y_1 - d_1}{\rho} & \left(-\frac{h_1}{2} \le y_1 < \frac{h_1}{2}\right) \\ \frac{(\frac{h_2}{2} - y_1)(\frac{h_1}{2} - d_1) - (\frac{h_2}{2} + y_1)(\frac{h_3}{2} + d_3)}{\rho h_2} & \left(-\frac{h_2}{2} \le y_2 < \frac{h_2}{2}\right) \\ \frac{y_3 - d_3}{\rho} & \left(-\frac{h_3}{2} \le y_3 \le \frac{h_3}{2}\right) \end{cases} \tag{2}$$

where $\rho$ represents the average radius of the multilayer. Results from the conventional bending theory (Eq. S2), FEM, and our new multi-neutral-axis model (Eq. 2) are compared in Fig. 2c. Conventional bending theory predicts monotonic linear strain distribution whereas both FEM and Eq. 2 capture non-monotonic strain distribution in the multilayer stack. Linear strain distribution in each layer obtained from FEM further corroborates our basic assumption. The locations of zero strain plane for different SU-8 thicknesses calculated using Eq. 1 are plotted in Fig. 2d. The dashed curve in Fig. 2d represents the equation $h_{3\mathrm{n}} = h_3$. When $h_3 < 8.28$ μm, the neutral axis will locate outside the SU-8 layer and hence no longer accessible for device placement. In conclusion, because of the soft silicone interlayer, the location of the neutral axis can be shifted away from near the center of the multilayer stack to within SU-8, a salient feature which enables photonic designs capitalizing on and enabling evanescent interactions while claiming extraordinary mechanical flexibility.

To experimentally validate the new multi-neutral-axis theory, we performed strain-optical coupling measurements, where the optical resonant wavelengths of glass micro-disk resonators were monitored *in-situ* while the samples were bent. A block diagram of the home-built measurement setup is illustrated in Fig. S4a. Light from a tunable laser was coupled into the bus waveguides via fiber end fire coupling, and the transmittance through the chip was monitored *in-situ* as the samples were bent using linear motion stages. Fig. 1d shows a flexible chip under testing on the setup. Further details regarding the measurement are provided in Supplementary Materials Part V.

The resonant wavelength shift $d\lambda$ can be expressed as a function of the local strain at the resonator $d\varepsilon$:

$$\frac{d\lambda}{d\varepsilon} = \sum_i \left[ \frac{\lambda}{n_g} \cdot \Gamma_i \cdot \left( \frac{dn}{d\varepsilon} \right)_i \right] + \frac{n_{eff}}{n_g} \cdot \frac{\lambda}{L} \cdot \frac{dL}{d\varepsilon} + \frac{\lambda}{n_g} \cdot \frac{dn_{eff}}{d\varepsilon} \tag{3}$$

where $\Gamma_i$ and $(dn/d\varepsilon)_i$ are the optical confinement factor and strain-optic coefficient in the *i* th cavity material, *L* is the cavity length, and $n_g$ and $n_{eff}$ denote the group index and effective indices, respectively. In Eq. 3, the first term on the right hand side (RHS) represents the optoelastic (i.e. strain-optic) *material* response, the second term manifests the cavity *length* change, and the third term results from the cavity *cross-sectional geometry* modification. Derivation of the equation follows Ref. 11 and is elaborated in Supplementary Material Part VI. Since the resonant wavelength of a high-Q resonator can be accurately measured down to the pico-meter level, strain-optical coupling provides a sensitive measure of local strain in the multilayer structure. A series of flexible $Ge_{23}Sb_7S_{70}$ glass micro-disk resonator samples with different SU-8 top and bottom cladding layer thickness combinations were fabricated and tested. By varying the cladding thickness, local strain at the micro-disk resonators is modified when the samples are bent. This is apparent from Fig. 3a, where the resonant wavelength shift as a function of chip bending curvature is plotted for 5 different samples. The resonant wavelength shift was highly repeatable after several bending cycles and little hysteresis was observed. Theoretical predictions based on the classical multilayer bending theory as well as those made using our new multi-neutral-axis analysis are plotted in the same figure for comparison. Experimentally measured material moduli and strain-optic coefficients measured using protocols outlined in Supplementary Materials Part IV were used in the calculations. It is apparent that the classical theory fails to reproduce the experimentally observed trend while our new theory successfully accounts for the strain-optical coupling behavior. When we apply Eq. 2 to convert the horizontal axis from bending curvature in Fig. 3a to bending-induced strain in Fig. 3b, all of the experimental data collapse onto one straight line, which verifies the linear dependence of resonance peak shift on mechanical strain (Eq. 3). The dramatic change of both magnitude and sign of the resonance shift in different samples provides an effective method to control strain-optical coupling in flexible photonic devices and also bears important practical implications: for applications where strain-optical coupling is undesirable such as resonator refractometry sensing, the coupling can be nullified by strategically placing the device at the zero strain points. On the other hand, the coupling can be maximized when applied to photonic tuning or strain sensing.

The local neutral axis design imparts extreme mechanical flexibility to our devices. To test the mechanical reliability of the flexible devices, micro-disk resonators were fabricated and their optical characteristics were measured after repeated bending cycles at 0.5 mm bending radius. As shown in Fig. 1f that there were minimal variations of both the quality factor and extinction ratio after multiple bending cycles, indicating superior mechanical robustness of the flexible devices. Optical microscopy further revealed no crack formation or interface delamination in the layers after the bending cycles.

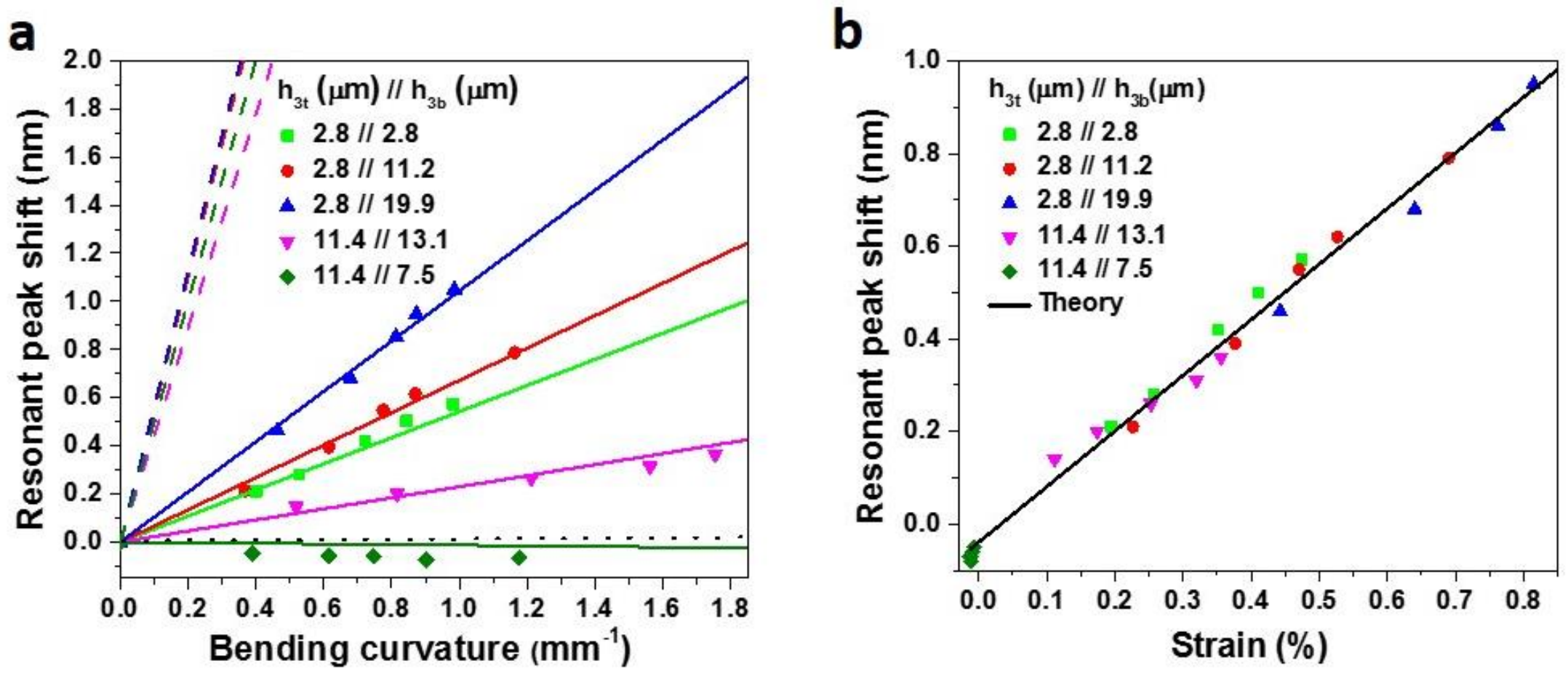

**Figure 3 | Strain-optical coupling in flexible photonic devices**. **a**, Resonance wavelength shift plotted as a function of bending curvature: each color represents an SU-8 top/bottom cladding thickness combination. $h_{3t}$ and $h_{3b}$ denote the SU-8 top and bottom cladding thicknesses, as labeled in Fig. 2a. The dots are experimentally measured data, the solid lines are predictions made using our analytical theory, and the dashed lines are classical bending theory results. **b**, Resonance wavelength shift plotted as a function of bending strain, which can be calculated from bending curvature using Eq. 2. All data in Fig. 3b collapse to one curve as predicted by Eq. 3.

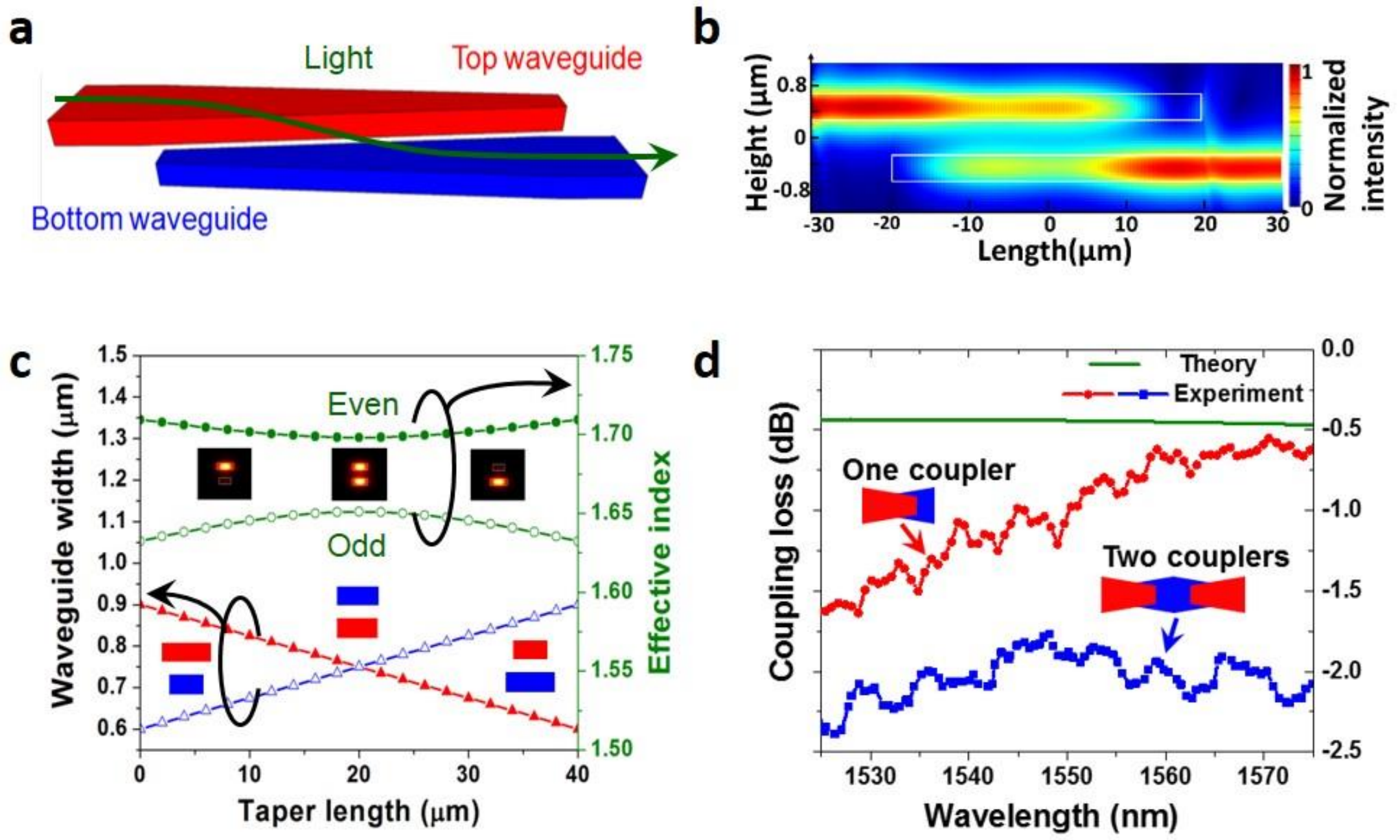

**Figure 4 | Adiabatic interlayer waveguide couplers.** **a**, Schematic structure of the interlayer waveguide coupler; **b**, Side view of steady-state optical field intensity distribution in the coupler, showing adiabatic power transfer from the top waveguide to the bottom one; **c**, Top (red curve) and bottom (blue curve) waveguide widths and simulated supermode effective indices (green curves) in the taper section of the interlayer coupler; the three insets show the cross-sectional even supermode intensity profile evolution along the taper; **d**, FDTD simulated (green line) and measured (red and blue lines) transmission spectra of the interlayer coupler(s).

Since our technology utilizes high-index ChGs as the backbone photonic materials, their amorphous nature further enables us to scale the fabrication method to 3-D monolithic photonic integration on plastic substrates using multilayer deposition and patterning. The excellent planarization capability of ultra-thin SU-8 resin ensures pattern fidelity in the multilayer process. This approach offers a facile and simple alternative for 3-D photonic structure fabrication to conventional methods involving ion implantation[16], wafer bonding[14], or pick-and-place nanomanipulation[15]. Here we demonstrate the fabrication of several important device building blocks including broadband interlayer waveguide couplers, vertically coupled resonators, and woodpile photonic crystals using our approach. It is worth noting that all devices presented in this paper were fabricated using simple, low-cost UV contact lithography without resorting to fine-line patterning tools such as electron beam lithography or deep-UV lithography, and we expect significant device performance improvement through further optimization of processing steps.

Fig. 4a schematically shows the structure of the interlayer adiabatic waveguide coupler. The coupler consists of a pair of vertically overlapping inverse taper structures made of $Ge_{23}Sb_7S_{70}$ glass. The non-tapered waveguide sections are 0.9 μm wide and 0.4 μm high, designed for single quasi-TE mode operation at 1550 nm wavelength. The device operates on the supermode adiabatic transformation principle[25,26], where light entering the coupler pre-dominantly remains in the coupled waveguide system's fundamental mode. The simulated field mode intensity profiles of the coupled waveguides in the taper section were plotted in Fig. 4c insets. The effective indices of the even mode and odd mode were also plotted along the taper length. As shown in the figure, the even supermode, which is the fundamental mode of the coupled waveguide system, transitions adiabatically from the top waveguide to the bottom waveguide as the waveguide width changes in the taper section. Fig. 4b shows a side view of finite-difference time domain (FDTD) simulated steady-state optical field intensity distribution in the coupler, which illustrates the power transfer process from the top waveguide to the bottom one. Unlike traditional directional couplers based on phase-matched evanescent coupling, the adiabatic mode transformer coupler design is robust against fabrication error and wavelength dispersion. The adiabatic interlayer coupler exhibited broadband operation with 1.1 dB (single coupler) and 2.0 dB (double couplers) insertion loss (both averaged over a 50 nm band), comparable to the simulation results (0.5 dB loss per coupler) given the limited alignment accuracy of contact lithography and waveguide sidewall roughness scattering loss (Fig. 4d).

Vertically-coupled resonator add-drop filters were fabricated using the same approach on plastic substrates (Fig. 5a). The device consists of a micro-disk resonator co-planar with the add waveguide, and a through-port waveguide in a second layer separated from the micro-disk by a 550 nm thick SU-8 layer. Unlike co-planar add-drop filters where coupling strength has to be adjusted by changing the narrow gap width between bus waveguides and the resonator, critical coupling regime in vertical resonators is readily achieved via fine tuning the SU-8 separation layer thickness. Fig. 5c shows the normalized transmission spectrum of a resonator designed for critical coupling operation. The filter exhibited an insertion loss of 1.2 dB and a loaded Q-factor of $2.5 \times 10^4$ at both it's through and drop port. These results agree well with our theoretical predictions made using a scattering matrix formalism[27] (Supplementary Materials Part VIII).

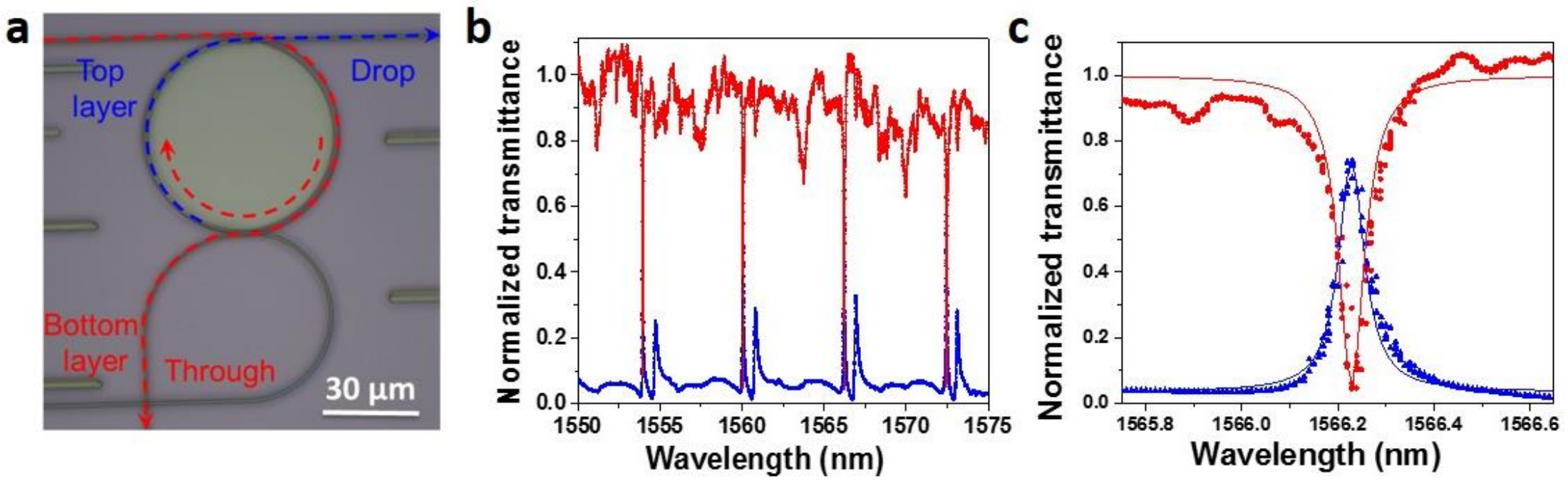

**Figure 5| Vertically coupled add-drop resonator filter**. **a**, Optical microscope image of a two-layer vertically coupled resonator. **b-c**, Normalized transmission spectra of a vertically coupled resonator at its through (red) and drop (blue) ports. The device is designed to operate at the critical coupling regime near 1550 nm wavelength. The dots represent experimental data and the lines are the theoretical results calculated using a scattering matrix formalism.

Besides two-layer devices such as interlayer couplers and vertical coupled resonators filters, our technique can be readily extended to the fabrication of multilayer structures which often present major challenges to conventional fabrication methods. As an example, Fig. 6a shows a tilted anatomy view of a four-layer woodpile photonic crystal (PhC) fabricated using the method shown in Fig. 1a prior to delamination from the handler substrate. The PhC structure integrity and pattern fidelity were examined using optical diffraction. Fig. 6b shows the diffraction spots from a collimated 532 nm green laser beam. The red dots in Fig. 6b represent the diffraction pattern calculated using Bragg equations (calculation details are presented in Supplementary Materials Part IX) which matches nicely with the experimental results. The well-defined diffraction pattern indicates excellent long-range structural order of the PhC.

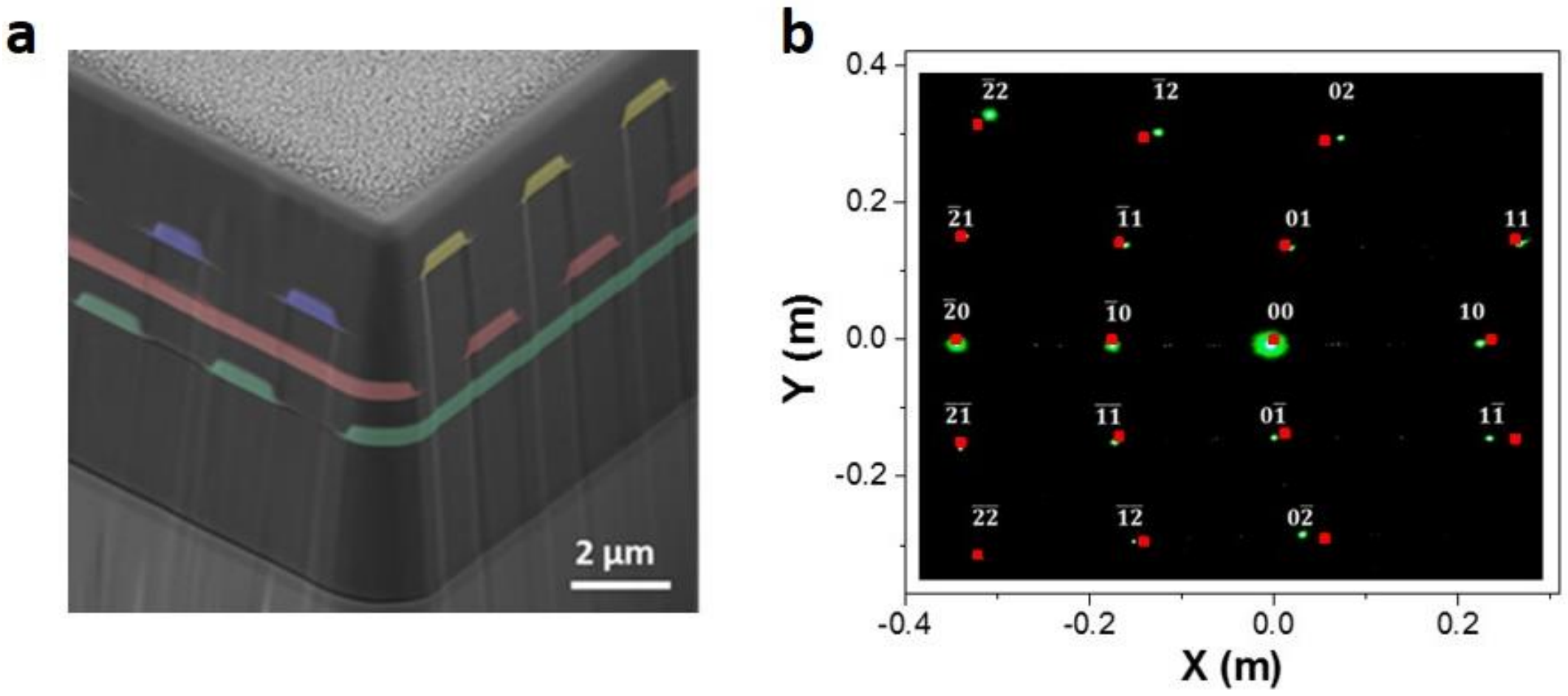

**Figure 6 | 3-D woodpile photonic crystals. a**, Tilted FIB-SEM view of a 3-D woodpile photonic crystal (prior to delamination from the Si handler substrate) showing excellent structural integrity. **b**, Diffraction patterns of a collimated 532 nm green laser beam from the PhC; the red dots are diffraction patterns simulated using the Bragg diffraction equation.

In conclusion, we have experimentally demonstrated a simple and versatile method to fabricate high-index-contrast 3-D photonic devices on flexible substrates. The method leverages the amorphous nature and low deposition temperature of novel ChG alloys to pioneer a 3-D multilayer monolithic integration approach with dramatically improved device performance, processing throughput and yield. A new nano-mechanical theory was developed and experimentally validated to accurately predict and control the strain-optical coupling mechanisms in the device. Guided by the nano-mechanical design theory, we demonstrated mechanically robust devices with extreme flexibility despite the inherent mechanical fragility of the glass film, and the devices can be twisted and bent to sub-millimeter radius without compromising their optical performance. The 3-D monolithic integration technique, which is applicable to photonic integration on both traditional rigid substrates and non-conventional plastic substrates, is expected to open up new application venues such as high bandwidth density optical interconnects, conformal wearable sensors, and ultra-sensitive strain gauges.

**Methods:**

**Material and Device Fabrication**

The device fabrication was carried out at the University of Delaware Nanofabrication Facility. First, an SU-8 epoxy layer was spin coated on the handler wafer. Then a negative photoresist (NR9-1000PY, Futurrex Inc.) pattern was lithographically defined on the SU-8 layer using contact lithography on an ABM Mask Aligner. ChG films were thermally evaporated onto the substrates from bulk glasses synthesized using a melt-quenching technique. The deposition was performed using a custom-designed system (PVD Products, Inc.). The deposition rate was monitored real-time using a quartz crystal micro-balance and was stabilized at 20 Å/s. After the deposition, the sample was sonicated in acetone to dissolve the resist layer, leaving a glass pattern reverse to that of the photoresist. The procedure is repeated several times to fabricate 3-D structures.

**Finite element simulations**

Finite element simulations were applied by ABAQUS 6.10 using plane strain elements (CPE4R) for the multilayer structure. In experiment, the concave bending of PI-silicone-SU-8 three-layer structure was actually induced through buckling mode (Fig. 1d) instead of pure bending. For this concave bending, PI layer which has larger bending rigidity compared with other two layers shares the most part of bending moment such that the bending of other two layers can be considered as being dragged by PI layer. We applied rotation boundary conditions on the two ends of PI layer which can generate curvatures the same with experiments at middle point of the structure to simulate this buckling induced bending.

**Optical transmission measurements**

The optical transmission spectra of waveguides and resonators in the C and L bands were collected using a fiber end-fire coupling approach shown in Fig. S4a. Tapered lens-tip fibers were used to couple light from an external cavity tunable laser (Agilent 81682A) into and out of the waveguides through end facets formed by cleaving the samples prior to delamination from the Si handler substrates. Fig. S4b shows a far-field image of quasi-TE guided mode output from a single-mode $Ge_{23}Sb_7S_{70}$ glass waveguide.

**Acknowledgements**

The authors thank S. Kozacik, M. Murakowski, and D. Prather for assistance with device fabrication, N. Nguyen and M. Mackay for mechanical tests, N. Xiao and Y. Liu for assistance with optical measurement data processing, and T. Gu and M. Haney for helpful discussions. L. L. acknowledges funding support from Delaware NASA/EPSCoR through the Research Infrastructure Development (RID) grant. H. L. and J. H. acknowledges funding support from National Science Foundation under award number 1200406. N. L. acknowledges the startup funding support from the Cockrell School of Engineering of UT Austin.

**Author Contributions**

L. L. and H. L. conducted material synthesis, optical modeling, device fabrication and testing. S. Q. and N. L. performed mechanics modeling and analysis. Y. Z. assisted in film depositions and device characterizations. J. H. conceived the device and structural designs. S. D., J. D. M., and K. R. contributed to material synthesis. J. H., N. L., and K. R. supervised and coordinated the project. All authors contributed to writing the paper.

**Supplementary Information**

# 3-D Integrated Flexible Glass Photonics

Lan Li[1†], Hongtao Lin[1†], Shutao Qiao[2†], Yi Zou[1], Sylvain Danto[3], Kathleen Richardson[3, 4], J. David Musgraves[4], Nanshu Lu[2], and Juejun Hu[1*]

*1. Department of Materials Science and Engineering*

*University of Delaware*

*Newark, Delaware 19716*

*2. Center for Mechanics of Solids, Structures and Materials*

*Department of Aerospace Engineering and Engineering Mechanics*

*University of Texas at Austin*

*Austin, Texas 78712*

*3. College of Optics and Photonics, CREOL*

*Department of Materials Science and Engineering*

*University of Central Florida*

*Orlando, Florida 32816*

*4. IRradiance Glass Inc.*

*Orlando, FL 32828, USA*

†These authors contributed equally to this work.

*e-mail: hujuejun@udel.edu

In this Supplementary Information, we provide further details on the device fabrication, characterization, and simulation results.

I. Ultra-thin SU-8 planarization tests

II. Propagating loss measurement of $Ge_{23}Sb_7S_{70}$ glass waveguides fabricated on flexible substrates

III. Mechanical measurements of material properties

IV. An analytical model accounting for multiple neutral axes in the multilayer stack

V. *In-situ* strain-optical coupling characterization method

VI. Strain-optical coupling theory

VII. Optimized 3-D photonic device design parameters

VIII. Scattering matrix formalism for calculating the transfer function of vertically coupled add-drop filters

IX. Woodpile photonic crystal diffraction experiment configuration

I. Ultra-thin SU-8 planarization tests

Planarization is the key to our multilayer 3-D fabrication process. We choose SU-8, an epoxy-based negative photoresist, as the planarizing agent. Unexposed SU-8 is a thermoplastic with a glass transition temperature ($T_g$) of 50 °C[1], which is amenable to thermal reflow processing at a relatively low temperature to create a smooth surface finish through the action of surface tension. Once cured or UV cross-linked, SU-8 transforms to a thermosetting polymer with a high $T_g$ > 200 °C and stable thermal/mechanical/chemical properties. To validate the planarization behavior of SU-8 on glass devices, a layer of 460 nm thick SU-8 epoxy was spin coated on gratings patterned in a 360 nm thick $Ge_{23}Sb_7S_{70}$ film (i.e. 100 nm over-coating layer thickness). Fig. S1a shows top-view optical micrographs of the grating patterns after SU-8 spin coating. The gratings have a fixed duty cycle of 0.5 and their periods are varied from 1.2 µm to 120 µm. The SU-8 layer topography after heat treatment was examined using cross-sectional SEM, and Fig. S1b (bottom) shows an exemplary SEM image of the grating structure after planarization. Transfer function of the planarization process was plotted as a function of spatial frequency (reciprocal of grating pitch) in Fig. S1c. Top panel of Fig. S1b schematically illustrated the definition of degree of planarization (DOP) and planarization angle (θ). The degree of planarization (DOP) is given by: $DOP = 100\% * \frac{t_1 - t_2}{t_2}$, and the planarization angle (θ) by $\theta = \arctan(\frac{t_1 - t_2}{W})$, where W is width of the grating line (i.e. half of the grating pitch). The ultra-thin (100 nm thickness) SU-8 overcoating layer produces DOP consistently larger than 98% for patterns with micron-sized pitch. Such planarization performance is considerably better than previous reports using BCB, PMMA or polyimide as the planarization agent, as shown by the comparison in Table S1.

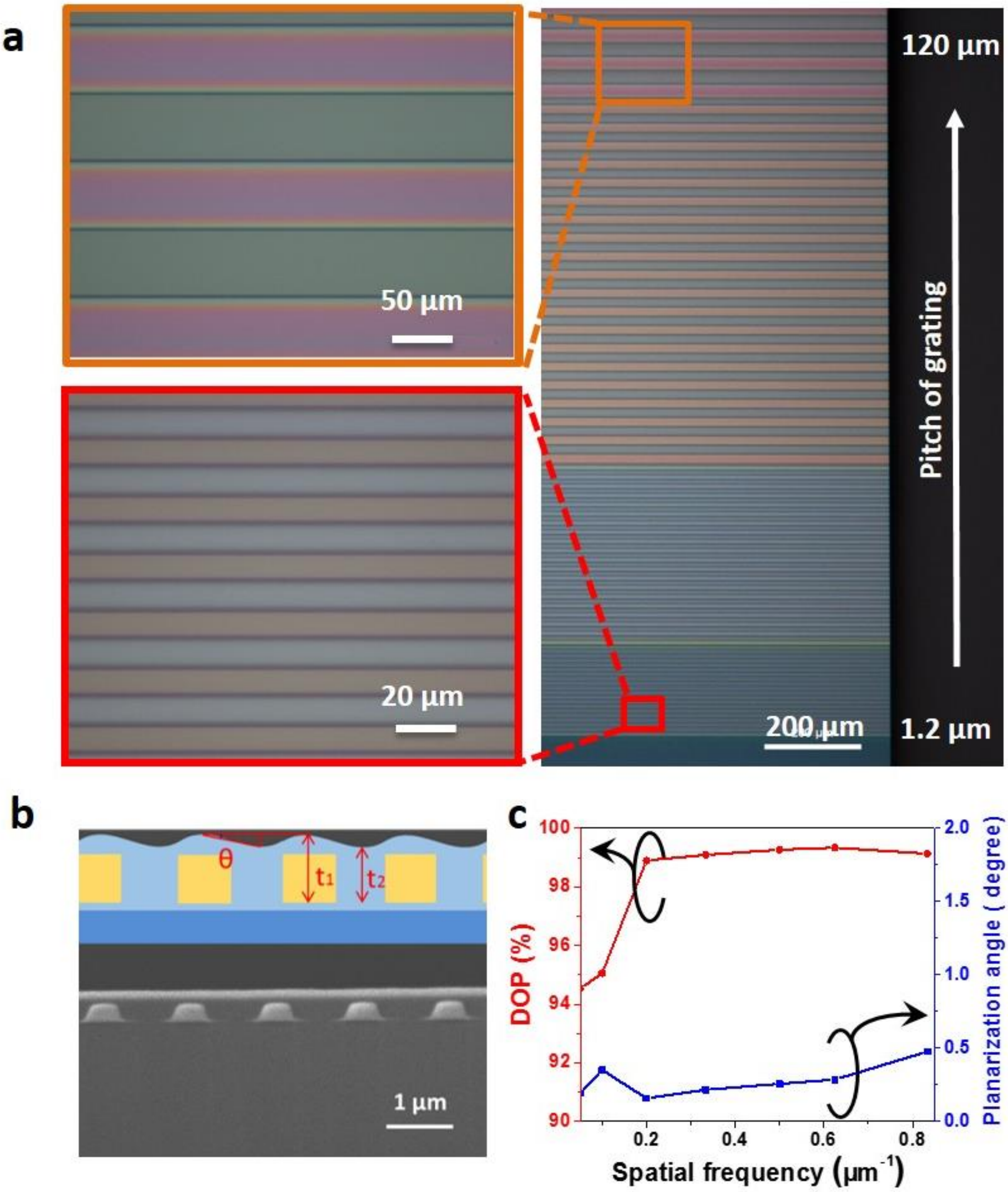

**Figure S1 | Ultra-thin SU-8 planarization characterizations**. **a**, Optical microscope images of glass grating patterns after single-layer SU-8 planarization. **b**, SEM cross-sectional image of planarized gratings; the top inset illustrates the definition of DOP and planarization angle. **c**, Plot of degree of planarization and planarization angle as functions of spatial frequency. The SU-8 planarization process consistently shows a DOP above 98% over micron-sized features and a small planarization angle < 1°.

**Table S1 |** Degree of planarization (DOP) of thin SU-8 compared to literature values using Polybenzocyclobutene (PBCB), Poly(methyl methacrylate) (PMMA) and polyimide (PI).

| Materials | Overcoating thickness ( µm) | Planarized topography | DOP |
|---|---|---|---|
| SU-8 (this report) | 0.1 | 5 µm pitch grating ( duty cycle 0.5) | 99% |
| | 0.1 | 20 µm pitch grating ( duty cycle 0.5) | 95% |
| BCB[2] | 0.3 | 10 µm wide trenches on a pitch of 30 µm | 90% |
| | 0.3 | 10 µm wide trenches on a pitch of 120 µm | 70% |
| PBCB[3] | 4.5 | 20 µm pitch grating ( duty cycle 0.5) | 85% |
| PI2555[3] | 1 | 20 µm pitch grating ( duty cycle 0.5) | 50% |
| PMMA[3] | 0.5 | 20 µm pitch grating ( duty cycle 0.5) | 18% |

II. Propagating loss measurement of $Ge_{23}Sb_7S_{70}$ glass waveguides fabricated on flexible substrates

Since the flexible substrates are not amenable to cleavage, it is difficult to use the cut-back method to measure waveguide propagating loss. Therefore, paper-clip waveguide structures (Fig. S2a) with different lengths and identical number of bends were fabricated for loss measurements using the end fire coupling method. Shown in Fig. S2b, a linear fit of the transmitted power as a function of waveguide length yields a propagating loss of 3.3 ± 0.6 dB/cm.

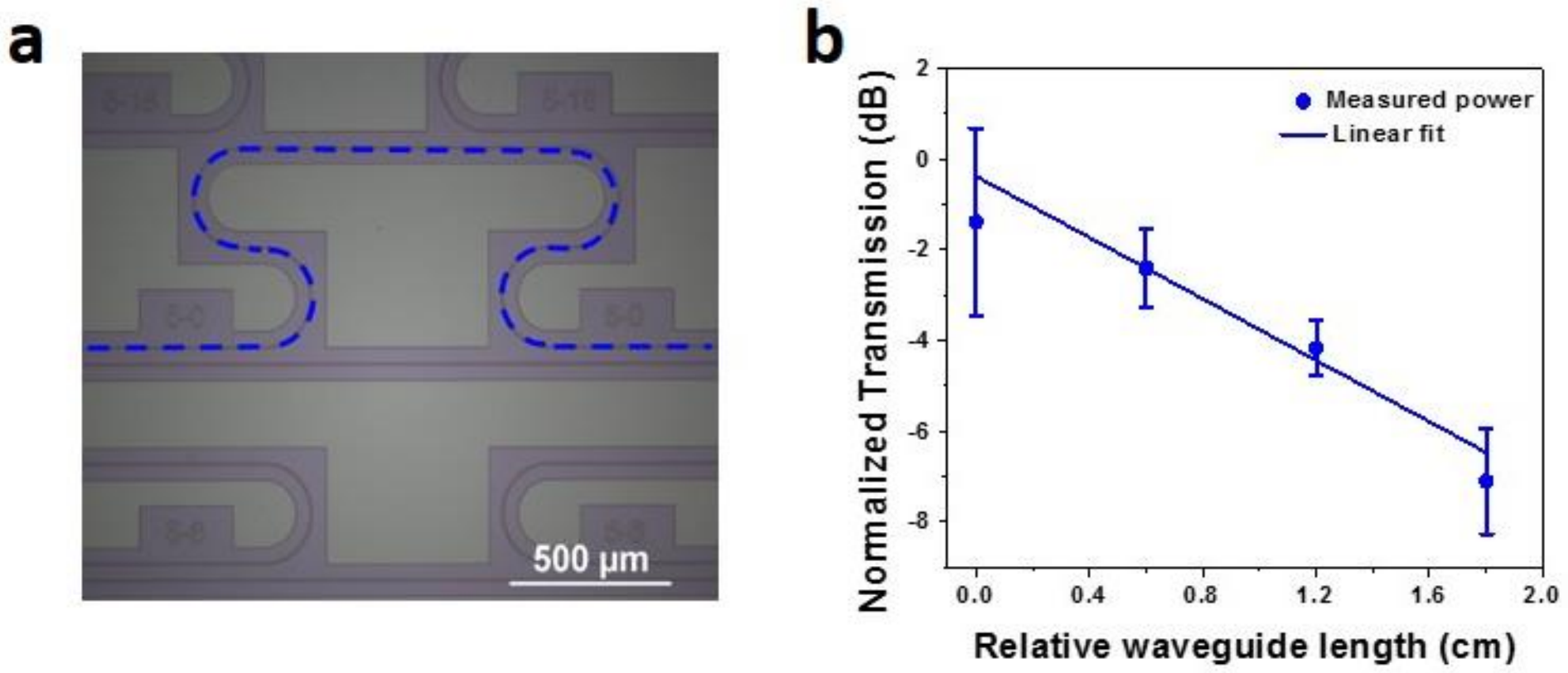

**Figure S2 | $Ge_{23}Sb_7S_{70}$ chalcogenide glass waveguides on flexible substrates**. **a**, Optical microscope image of paper-clip waveguide structures (the blue line). **b**, Transmitted optical power as a function of waveguide length at 1550 nm wavelength. The waveguide propagation loss is fitted from the graph to be 3.3 ± 0.6 dB/cm.

## III. Mechanical measurements of material properties

The Young's modulus of the constituent materials were experimentally extracted from uniaxial tensile tests performed on an RSA3 dynamic mechanical analyzer (TA instruments) with a Hencky strain rate of 0.01 $s^{-1}$. The layer thicknesses were measured by optical microscopy (Fig. S3a). The uniaxial strain-stress characteristics of the silicone adhesive layer and the polyimide layer were measured on the Kapton tapes. Specifically, two sets of experiments were performed: a double-layer Kapton tape was stretched in-plane to determine the modulus of polyimide, and a stack of Kapton tapes (Fig. S3a) underwent out-of-plane tensile tests to extract the strain-stress behavior of the silicone adhesive layer. In the former case, since the modulus of polyimide is over three orders of magnitude higher than that of silicone, the in-plane stress is pre-dominantly sustained by the polyimide layer. In the latter case, the strain primarily occurred in the silicone layer given its much lower modulus. The two measurements are analogous to the parallel and series connection of two elastic spring with vastly different spring constants (the inset of Fig. S3b-c). Both types of experiments were repeated multiple (> 6) times for statistical averaging. The statistically averaged strain-stress curves of polyimide and silicone calculated from the measurement results are plotted in Fig. S3b-c.

The strain-optical coefficients are essential parameters for calculating the strain-optical coupling strength in the flexible resonator devices, as suggested by Eq. 1. However, measuring strain-optical coefficients, in particular those of thin film materials (which can be significantly different from those of bulk materials for amorphous glasses), was regarded as extremely challenging task. Here we leverage the flexible resonator devices as a new measurement platform for accurate quantification of strain-optical coefficients of the materials. For resonators made of the same material and of the identical dimensions, Eq. 1 suggests the same $d\lambda/d\varepsilon$ coefficient regardless of the flexible device configuration. This conclusion is supported by our experimental results: if we re-plot the strain-induced resonant wavelength shift data (Fig. 3a) as a function of local strain at the resonators, it is clear that all data points fall on a single straight line (Fig. 3b), indicating the same $d\lambda/d\varepsilon$ coefficient (i.e. slope of the line). Since the second and third terms in Eq. 1 (resonator geometry change) can be readily inferred from our mechanical simulations, the material strain-optical response can be calculated by subtracting the contributions from the two terms. The calculation was used to generate the theoretical results plotted in Fig. 3a.

**Table S2 |** Materials parameters used in our FEM simulations

| Materials | Young's Modulus (MPa) | Density ( g/cc) | Poisson ratio |
|---|---|---|---|
| Polyimide | Fig. S3b | 1.42 | 0.34 |
| Silicone | Fig. S3c | 0.97 | 0.49 |
| SU-8 | 2000 | 1.12 | 0.22 |

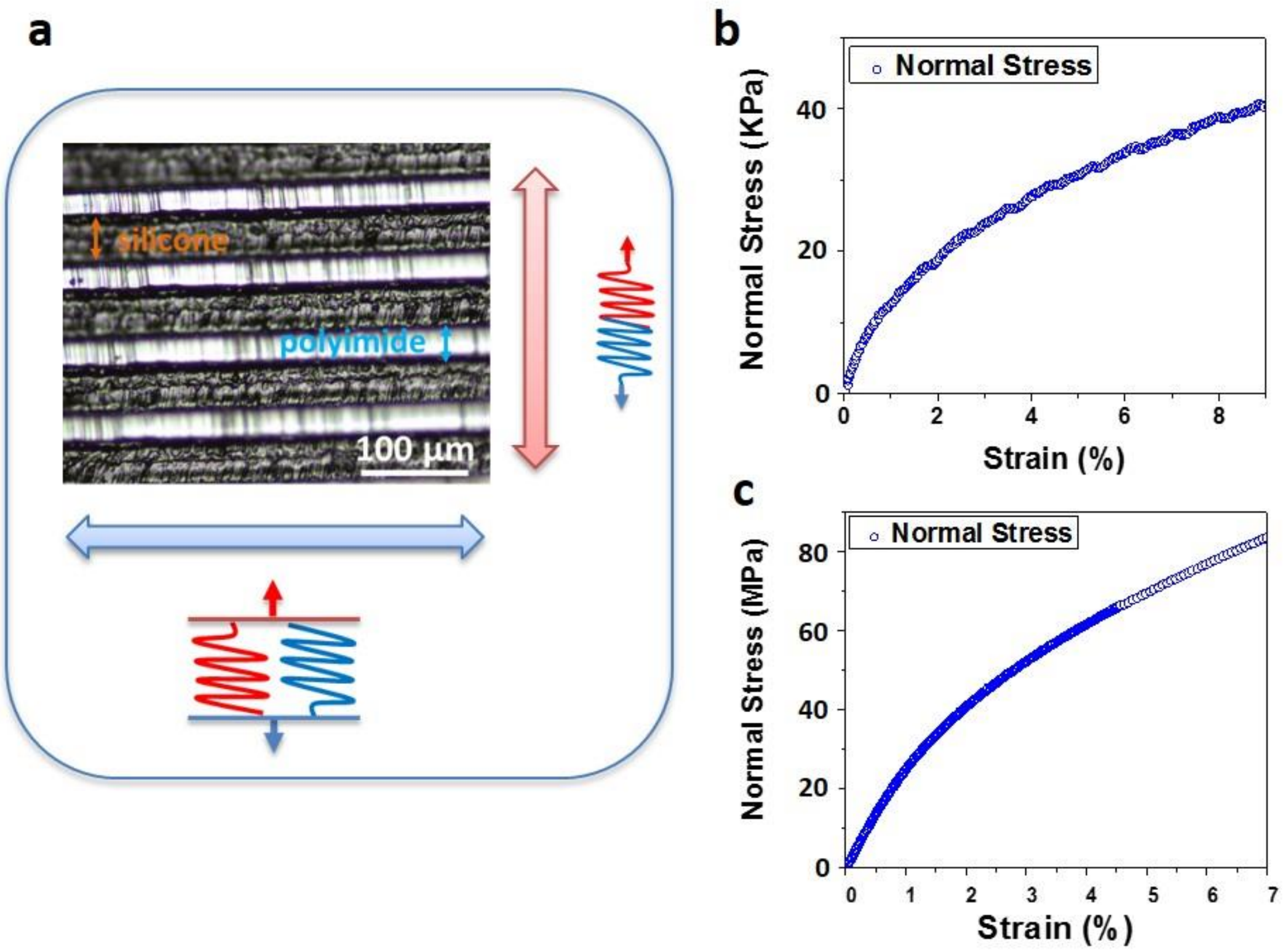

**Figure S3 | Mechanical tests of materials used in the flexible photonic device fabrication. a**, Cross-sectional optical microscope image of a Kapton tape stack. **b**, Typical stress-strain curve of the silicone layer. **c**, Typical stress-strain curve of the polyimide layer.

IV. An analytical model accounting for multiple neutral axes in the multilayer stack

When a multilayer structure is subject to pure bending, cross-sectional planes before bending are assumed to remain planar after bending in the classical bending theory. Under this assumption, a unique neutral axis of the laminated structure characterized by the distance *b* from the top surface of PI can be written as:

$$b = \frac{\sum_{i=1}^{3} \bar{E}_i h_i \left[ \left( \sum_{j=1}^{i} h_j \right) - \frac{h_i}{2} \right]}{\sum_{i=1}^{3} \bar{E}_i h_i} \qquad \text{(S1)}$$

where $\bar{E}_i = E_i / \left(1 - \nu_i^2\right)$ represents the plane strain Young's modulus with $\nu_i$ being the Poisson's ratio. Bending-induced tensile strain can be calculated analytically using:

$$\varepsilon = \frac{y}{\rho} \qquad \text{(S2)}$$

where *y* is the distance from the point of interest to the neutral axis and $\rho$ represents the radius of the neutral axis. Strain along the neutral axis is exactly zero and strains on different sides of the neutral axis have opposite signs.

Eq. S2 is only applicable to multilayer stacks with similar elastic stiffness. Since PI and SU-8 are much stiffer compared to silicone adhesive, we assume that inside each of the two layers, cross-sectional planes remain planar after bending so that linear relation between strain and curvature is still valid (Eq. S2).

The coordinate system we adopt is depicted in Fig. 2a, with notations given in Table S3.

**Table S3** | Notations used in the mechanical modeling

| Layer | Young's Modulus | Thickness | Strain distribution | Distance from neutral axis to median plane | Distance from neutral axis to bending center |
|---|---|---|---|---|---|
| PI | $\bar{E}_1$ | $h_1$ | $\epsilon_1(y_1)$ | $d_1$ | $\rho_1$ |
| silicone | $\bar{E}_2$ | $h_2$ | $\epsilon_2(y_2)$ | -- | -- |
| SU-8 | $\bar{E}_3$ | $h_3$ | $\epsilon_3(y_3)$ | $d_3$ | $\rho_3$ |

Our goal is to find an analytical solution for $d_3$ so that we know where to place the photonic devices for minimum strain.

Using the local coordinate systems, strain in PI can be written as:

$$\epsilon_1(y_1) = \frac{y_1 - d_1}{\rho_1} \; (-\frac{h_1}{2} < y_1 < \frac{h_1}{2}) \tag{S3}$$

Strain in SU-8 layer is

$$\epsilon_3(y_3) = \frac{y_3 - d_3}{\rho_3} \; (-\frac{h_3}{2} < y_3 < \frac{h_3}{2}) \tag{S4}$$

We suppose that strain in silicone layer is a linear function of $y_2$ as inspired by our FEM results:

$$\epsilon_2(y_2) = a y_2 + b \; (-\frac{h_2}{2} < y_2 < \frac{h_2}{2}) \tag{S5}$$

where $a$ and $b$ are coefficients to be determined by continuity conditions at PI/silicone and silicone/SU-8 interfaces.

Continuity at layer interfaces imposes:

$$\begin{cases} \epsilon_2\left(-\frac{h_2}{2}\right) = \epsilon_1\left(\frac{h_1}{2}\right) \\ \epsilon_2\left(\frac{h_2}{2}\right) = \epsilon_3\left(-\frac{h_3}{2}\right) \end{cases} \tag{S6}$$

Force equilibrium in *x* direction imposes:

$$\bar{E}_1 \int_{-\frac{h_1}{2}}^{\frac{h_1}{2}} \frac{y_1 - d_1}{\rho_1} dy + \bar{E}_3 \int_{-\frac{h_3}{2}}^{\frac{h_3}{2}} \frac{y_3 - d_3}{\rho_3} dy_3 + \bar{E}_2 \int_{-\frac{h_2}{2}}^{\frac{h_2}{2}} (a y_2 + b) dy_2 = 0 \quad \text{(S7)},$$

Combing Eqs. (S6) and (S7) yields a relation between $d_1$ and $d_3$:

$$2\left(\frac{\bar{E}_1 h_1 d_1}{\rho_1} + \frac{\bar{E}_3 h_3 d_3}{\rho_3}\right) = \bar{E}_2 h_2 \left(\frac{\frac{h_1}{2} - d_1}{\rho_1} - \frac{\frac{h_3}{2} + d_3}{\rho_3}\right) \tag{S8}$$

which satisfies the following two extreme cases: (i) When $E_2$=0, i.e. there is no material layer between PI and SU-8, both $d_1$ and $d_3$ should go to zero. (ii) When $\bar{E}_1 = \bar{E}_2 = \bar{E}_3$, i.e. the composite beam decays to a beam of uniform material, there should be only one neutral axis, i.e. $d_1 = \frac{h_2 + h_3}{2}$ and $d_3 = -\frac{h_1 + h_2}{2}$.

Assuming PI and SU-8 have the same original length and the same bending angle, the bending radius from each layer's neutral axis to its bending center should also be the same, which means

$$\rho_1 = \rho_3 \tag{S9}$$

Substituting Eq. (S9) into Eq. (S8) yields:

$$2(\bar{E}_1 h_1 d_1 + \bar{E}_3 h_3 d_3) = \bar{E}_2 h_2 \left((\frac{h_1}{2} - d_1) - (\frac{h_3}{2} + d_3)\right) \tag{S10}$$

As shown in Table S2, $\bar{E}_2 \ll \bar{E}_1$ and $\bar{E}_2 \ll \bar{E}_3$, which implies that the right hand side of Eq.

S10 is much smaller than the left hand side and can be approximated as zero, hence

$$d_3 = -d_1 \frac{\bar{E}_1 h_1}{\bar{E}_3 h_3} \quad \text{(S11)}$$

For a constant thickness of PI, our FEM results have validated a hypothesis that due to the soft silicone interlayer, the thickness of SU-8 has little effect on the position of the neutral axis of PI layer as long as SU-8 is thinner or of comparable thickness to the PI. We may then take $d_1$ as a constant and by fitting the FEM results, we obtain $d_1 = 0.836\ \mu\text{m}$. The distance from SU-8 surface to SU-8 neutral axis is finally given by Eq. 3 and plotted in Fig. 2d.

V. *In*-situ strain-optical coupling characterization

Fig. S4a schematically illustrates the flexible chip testing setup. Fig. S4b shows a far-field image of quasi-TE guided mode output from a single-mode $Ge_{23}Sb_7S_{70}$ glass waveguide on a plastic substrate and Fig. S4c shows a photo of a flexible chip under bending test. The chip was glued onto the linear motion sample stages using double-sided sticky tapes. Bending radius of the flexible chip was measured from the image using an imaging processing software (Image J).

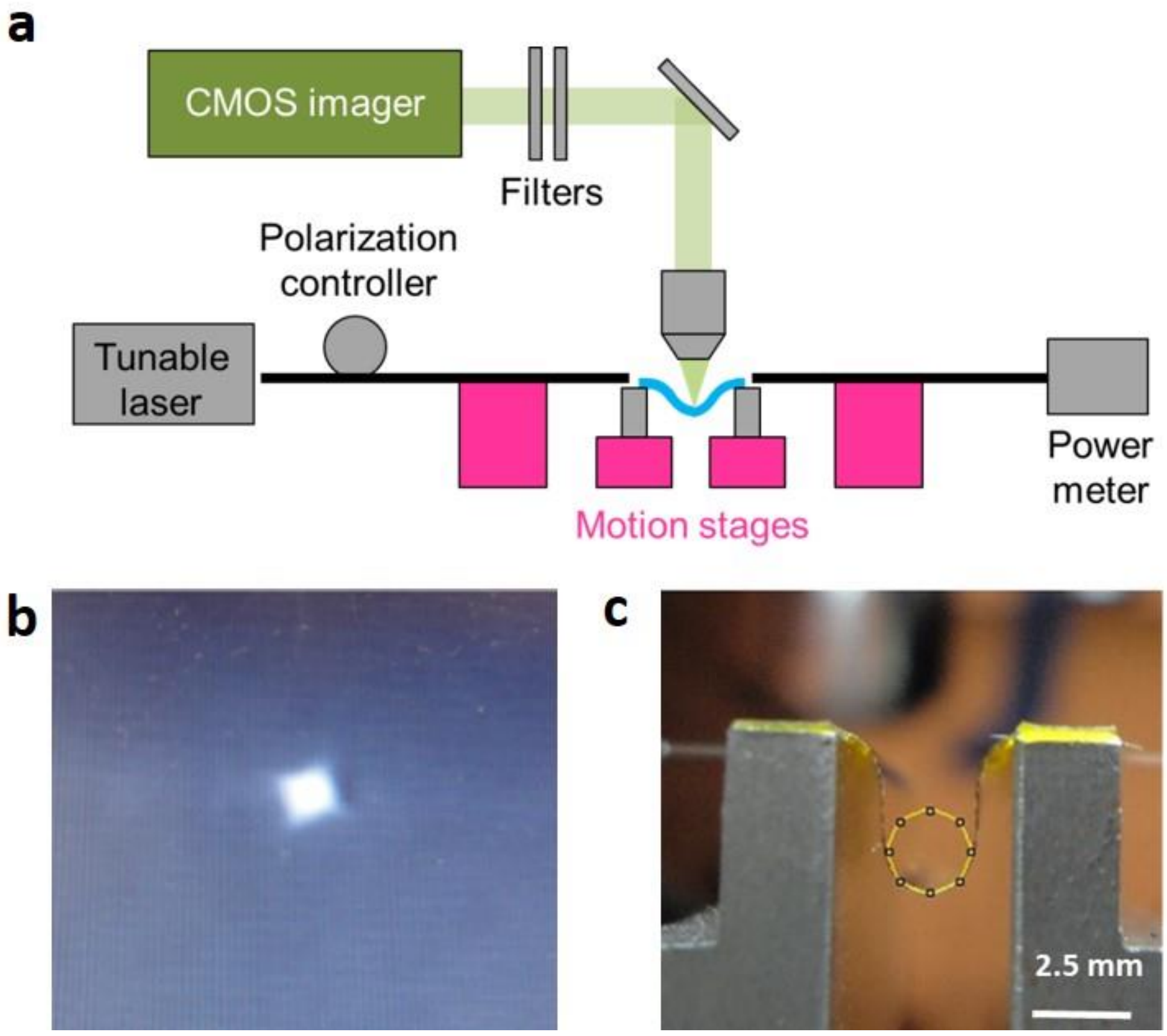

**Figure S4 | a**, Schematic diagram of the testing setup. **b**, Far-field image of TE guided mode output from a single-mode flexible $Ge_{23}Sb_7S_{70}$ glass waveguide. **c**, Photo of a flexible waveguide chip under bending test: an imaging processing software was used to extract the bending radius of the chip from the image.

## VI. Strain-optical coupling theory

This Part of Supplementary Material aims to derive Eq. 3, which governs the strain-optical coupling properties in flexible photonic resonant cavity devices.

The resonant condition of a resonant cavity device can be generally given by:

$$\mathrm{N}\lambda = n_{eff}L \tag{S12}$$

when strain introduces a perturbation to the resonant wavelength λ, we have:

$$\mathrm{N}(\lambda + \mathrm{d}\lambda) = n_{eff}L + d(n_{eff}L) = n_{eff}L + Ldn_{eff} + n_{eff}dL \tag{S13}$$

The effective index change $dn_{eff}$ originates from two mechanisms: the strain-optical material response, as well as the cross-sectional geometry change of the resonator due to strain:

$$dn_{eff} = \sum_i \Gamma_i \left(\frac{dn}{d\varepsilon}\right)_i d\varepsilon + \frac{dn_{eff}}{d\varepsilon}d\varepsilon + \frac{dn_{eff}}{d\lambda}d\lambda \tag{S14}$$

Here the first term corresponds to the material contribution, and the second term only takes into account the geometric change. The coefficient $dn_{eff}/d\varepsilon$ can be calculated once the local strain at the resonator is known using the perturbation theory involving shifting material boundaries[4]. The third term results from effective index dispersion in the resonator: as the resonant wavelength shifts, the corresponding effective index also changes. Combining the two equations leads to Eq. 3. Note that the group index in Eq. 3 is defined as:

$$dn_{eff} = n_{eff} - d\lambda\frac{dn_{eff}}{d\lambda} \tag{S15}$$

VII. Optimized 3-D photonic device design parameters

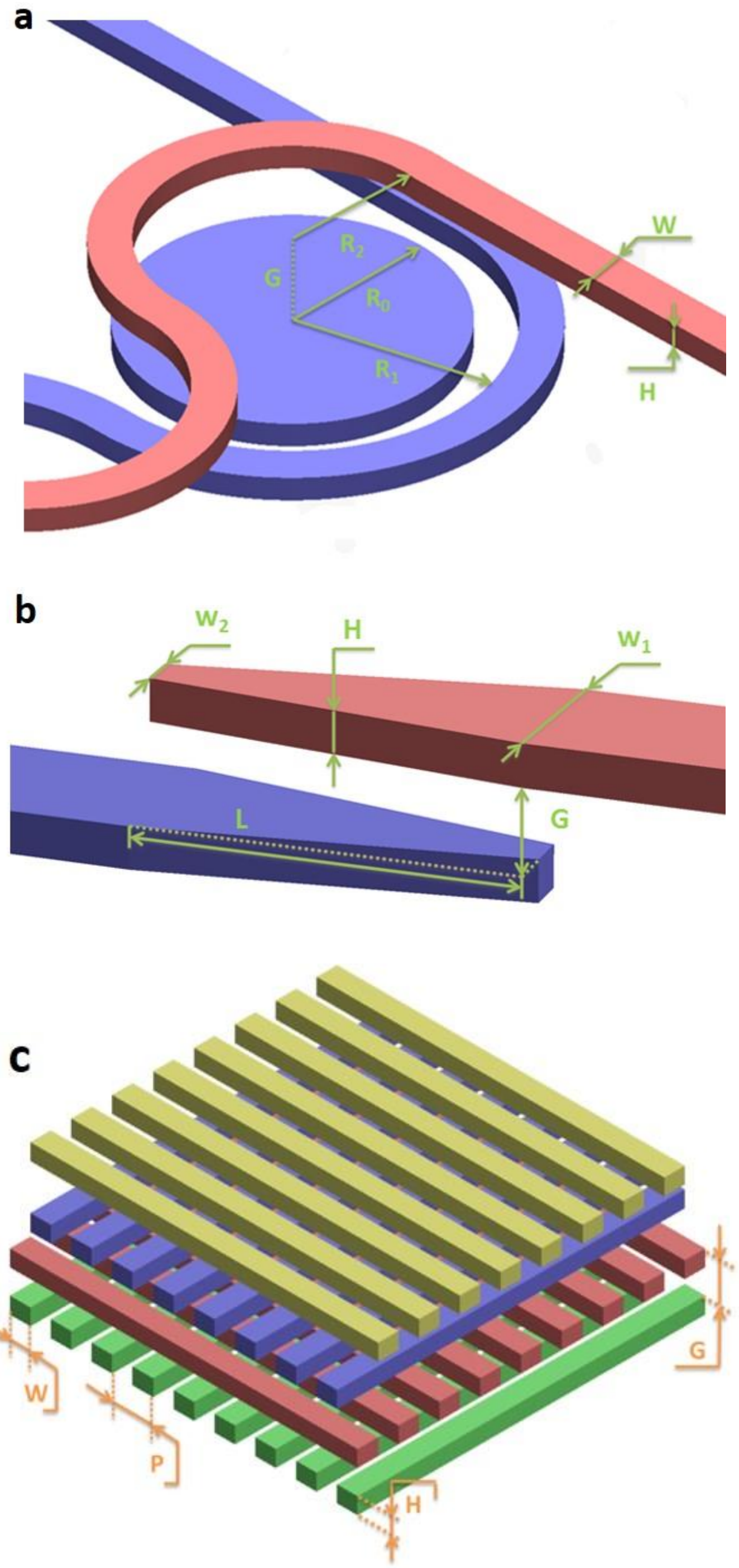

**Figure S5 | Schematic structures of 3-D photonic devices. a**, a two-layer vertically coupled resonator. **b**, An interlayer waveguide couplers. **c**, A woodpile photonic crystal. The key geometric parameters are labeled in the figures.

**Table S5 |** Optimized 3-D photonic device geometric parameters.

| | | | | | | |
|---|---|---|---|---|---|---|
| Fig. S5a | W | H | $R_0$ | $R_1$ | $R_2$ | G |
| (μm) | 0.8 | 0.4 | 30 | 30.8 | 29.4 | 0.9 |
| Fig. S5b | $W_1$ | $W_2$ | L | H | G | |
| (μm) | 0.9 | 0.6 | 40 | 0.4 | 0.55 | |
| Fig. S5c | W | P | H | G | | |
| (μm) | 1.5 | 3 | 0.36 | 0.8 | | |

A set of 3-D photonic devices with varied design parameters were fabricated and tested. Table S5 summarizes the experimentally determined geometric parameter combinations that yield the optimized performance presented in this study.

## VIII. Scattering matrix formalism for calculating the transfer function of vertically coupled add-drop filters

The vertically coupled add-drop filter can be treated using a circuit model[5] shown in Fig. S6. The power coupling coefficients between the resonator and the add and drop port waveguides are given by $K_1$ and $K_2$, respectively. Power loss due to coupling is neglected in the calculation. L is the resonator circumference and α is the optical loss in the micro-disk resonator. Using the scattering matrix method, loaded quality factor $Q_{load}$, maximum power at the drop port and minimum power at the through port at resonance can be obtained by the equations:

$$Q_{load} = \frac{2\,\pi n_g L}{\lambda_r(K_1+K_2+(1-\sigma^2))} \tag{S16}$$

$$I_{drop-max} = I_{in}\frac{K_1K_2\sigma^2}{(1-\sigma\sqrt{1-K_1}\sqrt{1-K_2})^2} \tag{S17}$$

$$I_{through-min} = I_{in}\frac{(\sqrt{1-K_1}-\sigma\sqrt{1-K_2})^2}{(1-\sigma\sqrt{1-K_1}\sqrt{1-K_2})^2} \tag{S18}$$

where λ is the resonant wavelength, $I_{in}$ is the input power, σ is the intrinsic amplitude loss in one round trip and is expressed as

$$\sigma = \exp(-1/2\ \alpha L) \tag{S19}$$

$n_g$ is the group index of the resonator and can be defined as follows:

$$n_g = c_0/(FSR \cdot L) \tag{S20}$$

Here, $c_0$ is the light speed in vacuum, and FSR is the free spectral range (FSR) of the resonator in the frequency domain. Based on the measured transmission spectrum (Fig. 5b) and the equations above, the coupling coefficients can be fitted, which in turn allows the prediction of the resonator transmission spectrum. The parameters used in the calculation are listed in Table S6.

The intrinsic quality factor ($Q_{in}$) of the micro-disk resonator is $1.9 \times 10^5$ calculated using[6]

$$Q_{in} = \frac{2\pi n_g}{\alpha\lambda} \tag{S21}$$

which is consistent with that of all-pass resonators we fabricated using the same method.

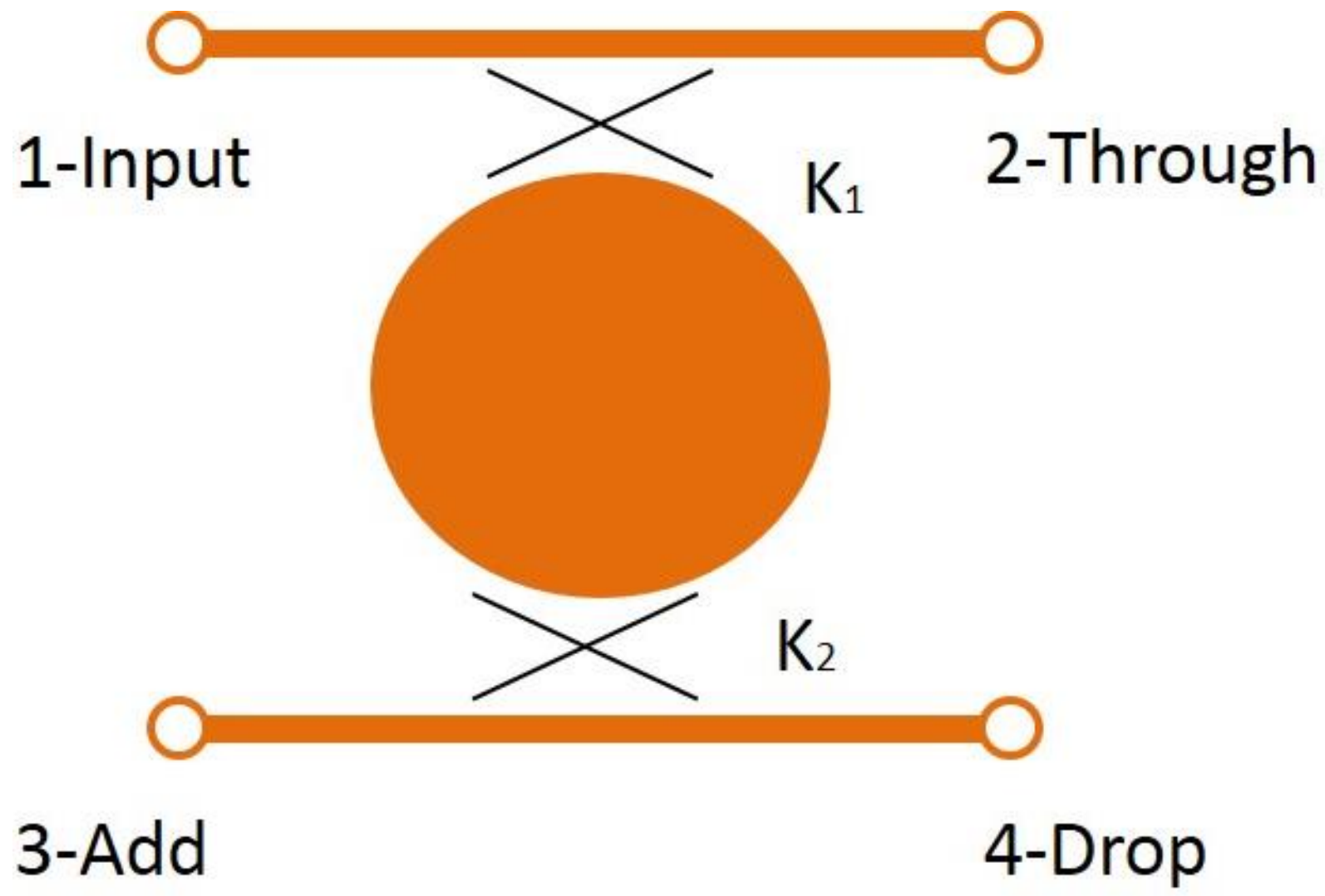

**Figure S6 |** Schematic illustration of a circuit model of the vertically coupled add/drop filter.

**Table S6 |** Parameters used to calculate the transmission spectrum of the vertically coupled add/drop resonator filter shown in Fig. 5c.

| L(um) | $\lambda_r$(nm) | $I_{in}$ (nW) | $I_0$ (nW) | $I_{thmin}$ (nW) | $I_{drmax}$ (nW) | $Q_{load}$ |
|---|---|---|---|---|---|---|
| 188.5 | 1566.228 | 87.7 | 3.2 | 4.4 | 67.3 | $2.2 \times 10^4$ |
| FSR(GHZ) | $n_g$ | $K_1$ | $K_2$ | σ | α(dB/cm) | $Q_{in}$ |
| 757 | 2.10 | 0.041 | 0.024 | 0.996 | 2.0 | $1.9 \times 10^5$ |

## IX. Woodpile photonic crystal diffraction experiment

Fig. S7 illustrates the diffraction experiment setup. Monochromatic green light from a collimated 532 nm diode-pumped solid state green laser was incident at an angle θ = 45 degrees upon the surface of the PhC sample, and the diffraction patterns were projected onto a white board perpendicular to the (0 0) specular order reflection. To calculate the projected positions of different diffraction orders, the wave vector can be decomposed in the coordinate system shown in Fig. S7, where the x and z axes are in-plane with the woodpile PhC structure while the y axis is perpendicular to the PhC sample surface. For the (0 0) specular order, the wave vector is:

$$\overleftarrow{k}_{(0\,0)} = \left(k_{x0}, k_{y0}, 0\right) = (k_0 cos\theta, k_0 sin\theta, 0) \quad \text{(S22)}$$

where $k_0$ is the wave vector in vacuum. For a diffraction spot of the (n m) order, its wave vector is:

$$\overleftarrow{k}_{(n\,m)} = \left(k_x, k_y, k_z\right) = (k_{x0} + \mathrm{n}\, G_x, \sqrt{k_0^2 - k_x^2 - k_z^2}, \mathrm{m}\, G_z\, ) \quad \text{(S23)}$$

according to the Bragg diffraction equation, where $G_x$, $G_z$ are the PhC in-plane reciprocal lattice vectors. Thus the diffraction angle of the (n m) order light with respect to the (0 0) order can be calculated using:

$$\alpha = \arctan\left(\frac{k_x}{k_y}\right) - \theta \quad \text{(S24)}$$

$$\beta = \arctan\left(\frac{k_z}{\sqrt{k_x^2 + k_y^2}}\right) \quad \text{(S25)}$$

and the position of the (n m) order pattern in the x'-y' plane (on the white board) is written as:

$$(\mathrm{D} \tan\alpha\ , \mathrm{D}\ \tan\beta\ / \cos\alpha)$$

Diffraction patterns predicted using the approach agree well with the experimental results, as shown in Fig. 6b.

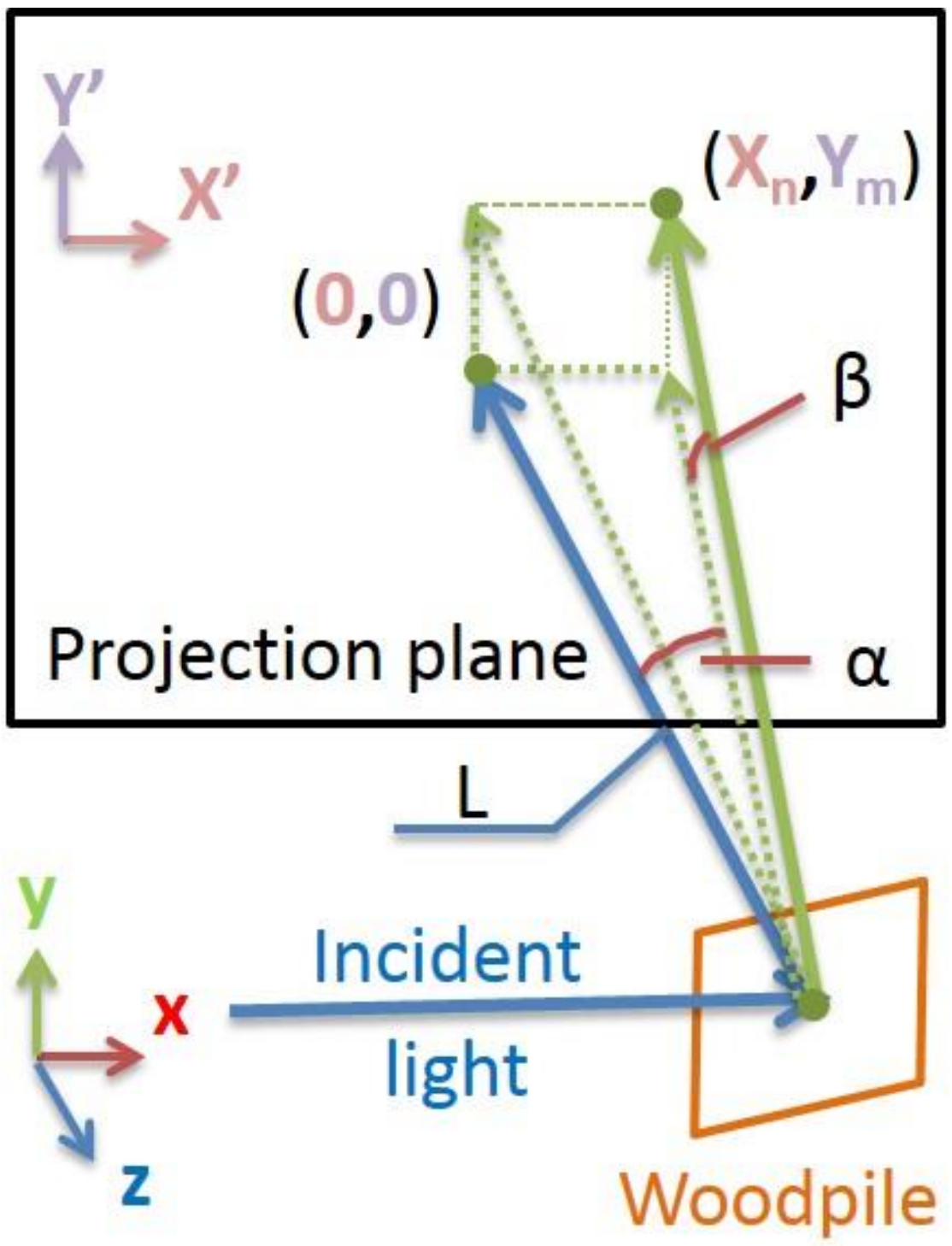

**Figure S7 |** Schematic diagram illustrating the experimental setup used to map the diffracting patterns from the woodpile photonic crystal structure.